\documentclass[11pt]{article}
\usepackage{amsmath, amsthm, amsfonts}
\usepackage[margin=1in,footskip=0.25in]{geometry}
\usepackage{mathtools}
\usepackage{makecell}
\usepackage{multicol}
\usepackage{graphicx}
\usepackage{hyperref}
\usepackage{cite}
\theoremstyle{theorem}
\usepackage{lscape}

\newtheoremstyle{defi}
  {10pt}          
  {10pt}  
  {\rm}  
  {\parindent}     
  {\bf}  
  {. }    
  { }    
  {}     
\theoremstyle{defi}

\begin{document}

\date{}

\title{\bf Qualitative behavior of  cosmological models combining various matter fields}
\author{Parth Shah$^{1}$, Gauranga C. Samanta$^{2}$, Salvatore Capozziello$^{3, 4, 5, 6}$ \\
 $^{1, 2}$Department of Mathematics,
BITS Pilani K K Birla Goa Campus,
Goa-403726, India,\\
 $^{3}$Dipartimento di Fisica "E. Pancini", Universit\'a di Napoli "Federico II", and
 $^{4}$INFN Sez. di Napoli, \\
 Compl. Univ. di Monte S. Angelo, Edificio G, Via Cinthia, I-80126,
Napoli, Italy. \\
 $^{5}$Gran Sasso Science Institute, Via F. Crispi 7, I-67100, L' Aquila, Italy,\\
$^{6}$ Tomsk State Pedagogical University, ul. Kievskaya, 60, 634061 Tomsk, Russia.\\
parthshah2908@gmail.com \\
gauranga81@gmail.com \\
capozziello@na.infn.it}

\maketitle

\begin{abstract} The late time accelerated expansion of the universe can be realized using
scalar fields with  given self-interacting potentials. Here we consider a straightforward approach where
 a three cosmic fluid mixture is assumed.  The fluids are  standard matter perfect fluid,
dark matter,  and  a scalar field with the role of dark energy. A dynamical system analysis is developed in this context.
A central role  is played by the equation of state $\omega_{eff}$ which determines  the acceleration
phase of the models. Determining  the domination of a particular fluid at certain stages of the
universe history by stability analysis allows, in principle, to establish the succession of the various cosmological eras.
\end{abstract}



\section{Introduction}
From the recent observational data, it is assumed that, our universe is spatially flat and suffered two acceleration phases. The first acceleration, occurred
at the time of the big bang, is called  inflation. It occurred before the radiation dominated era. The second accelerated phase occurred after the matter dominated era and it is lasting up to now. Since the discovery of this late accelerated expansion \cite{riess, perlmutter}, there have been several studies  searching for  candidates capable of sourcing such a dynamics. First, and foremost
among them, is  assuming a positive cosmological constant $\Lambda$ in the framework of the so-called $\Lambda$CDM where cosmological constant and cold dark matter supply the largest part of the matter-energy content of the universe. Even though this model fits very well with recent astronomical observational data,  there are some unresolved issues,  the so-called {\it cosmological constant problem} \cite{weinberg, martin} and {\it coincidence problem} \cite{zlatev}, related to the tiny value of $\Lambda$ and its fine tuning with today observed matter content of the universe. Another approach to handle the problem is considering the observed value of cosmological constant to be dynamically derived from some evolutionary process. This result can be achieved by introducing some  scalar field which evolves giving rise to  the late time acceleration of the universe. Any entity which, at  cosmic distances, provides an accelerated expansion in late epoch is generally  termed as {\it dark energy}. We have two basic approaches for explaining dark energy. The first way is to ``modify matter" i.e. the $T_{\mu\nu}$  stress-energy tensor   in Einstein's field equation which would contain some form of matter-energy  with negative pressure. The second way is to ``modify geometry"   leading to changes in the Einstein field equations  ${\displaystyle G_{\mu\nu}=R_{\mu\nu}-\frac{1}{2}g_{\mu\nu}R}$ (see  for example \cite{Amendola}).

Here we are considering modifications  of matter content  by adding terms with negative pressure. This  purpose  is fulfilled by adding a generic  scalar field which is one of the components of the cosmic fluid. Our aim is discussing a cosmological  qualitative behavior by which late acceleration is generically achieved by summing up suitable fluids.
With this perspective in mind,
dark energy equation of state parameter $\omega$ between $-\frac{1}{3}$ and $-1$ is termed as quintessence \cite{copeland, tsujikawa}. Scalar fields are very important in cosmology not just because they represent dark energy behavior, but  also because, at different scales, they give rise to inflationary behavior, dark matter models and other cosmological features \cite{Arturo}.

One of the major problems in theories of gravity is the difficulty in finding out  (analytic or numerical) solutions due to highly non-linear terms in the field equations and hence comparison with observations cannot be carried out easily. So it is important that other techniques are efficiently used to solve such equations or, at least, to control the overall dynamical behavior. A method is the Dynamical System Analysis. Application of dynamical systems analysis to cosmology has been widely discussed in literature \cite{Ellis,Coley}. In general dynamical systems are adopted to solve several problems in physics \cite{Place, Hirsch, Lefschetz, Lynch, Perko, Wiggins}\bibliographystyle{IEEEtran}.  The approach  is a method for finding out  numerical solutions and  helps in  understanding the qualitative behavior of a given physical  system \cite{Roy, Kbamba, Hohmann, Carneiro, Bhatia, Wainwright, Copeland1, Santos, Bamba, Odintsov, Odintsov1}.
The most important concept in dynamical system analysis  is to find out  critical points of a set of  first order ordinary differential equations (ODE). They are the zeros of a vector field $f(x,y,z,...)$. Phase space stability  points are those that satisfy the condition $x' = 0, y' = 0, z' = 0,..$ where the prime means derivative with respect the affine parameter (time in particular). The stability conditions are obtained by calculating the Jacobian matrix at critical points and finding their eigenvalues. After the identification of critical points and eigenvalues we can identify the flow near any of these points by linearizing the system in a neighborhood of the point. The fundamental idea is to analyze if  trajectories in the neighborhood of the point are either attracted or repelled. This is the study of stability properties near a particular critical point.

In the present work we  analyze the stability of cosmological models assumed to be sourced by  various kinds of fluids. We start our analysis  considering a completely dark matter filled universe, Subsequently, we assume the   universe completely filled by a  standard perfect fluid (with equation of state $p = \omega \rho$). Then we consider  non-interacting mixtures of various fluids assuming dark matter together with perfect fluid. Then, we introduce dark energy and consider the universe filled by a mixture of dark matter, perfect fluid and dark energy.  We consider the various components  and discuss the stability of the cosmological model accordingly. Finally, we consider a mixture of dark matter, perfect fluid and scalar field with a given potential as a candidate for dark energy. Specifically we consider  exponential and power law potential. In each  case, we  analyze  stability conditions related to  $\omega_{eff}$ as well as the acceleration phase.
The paper layout is the following. In Sec. 2, we discuss the various fluids sourcing the universe that we are going to analyze. Dynamical system analysis is performed in Sec.3. Here we take into account the various combinations of the given fluids and study the critical points accordingly. Conclusions and perspectives are reported in Sec.4.

\section{Cosmic fluids}
The generic action of General  Relativity where geometry is minimally coupled to perfect fluid matter and scalar field can be written as
\begin{equation}
\label{GR}
S = \int d^4x \sqrt{-g} \left[ R + {\cal L}_m + \cal L_{\phi} \right]
\end{equation}
where $R$ is Ricci Scalar, $g$ is the metric determinant, ${\cal L}_m$ is the matter Lagrangian, and  ${\cal L_{\phi}} = \frac{1}{2} \partial \phi^2 - V(\phi) $ is the scalar field Lagrangian. The metric is $g_{\mu\nu}$ = diag$(-1,a^2(t),a^2(t),a^2(t))$ and $a(t)$ is scale factor of the universe.

Let us consider the  source  divided into three parts, i.e.  Dark Matter (DM), Dark Energy (DE) and standard matter (perfect fluid PF).

Dark matter is assumed to be described as dust $(p_{DM}=0)$.  Its energy-momentum tensor can be defined as

\begin{equation}
  T_{\mu\nu}^{DM}=\rho_{DM}u_{\mu}u_{\nu}
\end{equation}
where $u^{\mu}=(1, 0, 0, 0)$ is a four velocity vector.

Dark energy is described by a scalar field $(\phi)$ (quintessence or phantom) rolling down a potential $V(\phi)$. The energy momentum tensor  is defined as
\begin{equation}\label{}
 T_{\mu\nu}^{\phi}=\left(\frac{1}{2} \epsilon \dot{\phi}^2+V(\phi)\right)u_{\mu}u_{\nu}+ \left(\frac{1}{2} \epsilon \dot{\phi}^2-V(\phi)\right)h_{\mu\nu}
\end{equation}
where $h_{\mu\nu}=g_{\mu\nu}+u_{\mu}u_{\nu}$.
The energy density $\rho_{\phi}$ and the isotropic pressure $p_{\phi}$ of the field are
\begin{equation}
\rho_{\phi} = \frac{1}{2} \dot{\phi}^2 + V(\phi)
\hspace{0.3in} \text{and} \hspace{0.3in}
p_{\phi} = \frac{1}{2} \dot{\phi}^2 - V(\phi),
\end{equation}
and the equation of state parameter can be written  as
\begin{equation}
\omega_{\phi} = \frac{p_{\phi}}{\rho_{\phi}} = \frac{\frac{1}{2} \dot{\phi}^2 - V(\phi)}{\frac{1}{2} \dot{\phi}^2 +V(\phi)}.
\end{equation}

Here, standard matter is assumed to be described by a perfect fluid with energy momentum tensor
\begin{equation}\label{}
  T_{\mu\nu}^{PF}=(p_{PF}+\rho_{PF})u_{\mu}u_{\nu}+p_{PF}g_{\mu\nu}
\end{equation}
and a linear equation of state
\begin{equation}\label{}
  p_{PF}=\omega\rho_{PF}, ~~ 0<\omega\le 1,
\end{equation}
in the so called Zeldovich interval. For $\omega=1 ~\& ~\frac{1}{3}$, it represents  stiff matter and radiation fluid model respectively. Actually, also $\omega=0$ can be inserted in the Zeldovich interval. In such a case, we are dealing with standard baryonic matter that, according to observations, represents a small percentage ($\leq 5\%$) of the total matter amount of the universe. For our dynamical analysis, we can assume that the bulk of matter content is represented by dark matter.
Throughout this work we will assume physical units, that is $\displaystyle \frac{8 \pi G}{c^4} = 1$.

\section{Dynamical system analysis}
With the above considerations in mind, let us study the dynamical systems for  the various combinations of fluids sourcing a Friedman-Robertson-Walker universe derived from the action \eqref{GR}. The aim is to study the stability of critical points  in view of deriving the overall behavior of the related cosmological model.
\\
\\
\textbf{Case-I:} Dark Matter $(p_{DM}=0)$
\\
\\
In this case, we have considered the universe filled with dark matter only. The explicit form of the field equations is:
\begin{equation}\label{8}
  3\left(\frac{\dot{a}}{a}\right)^2=\rho_{DM}
\end{equation}
\begin{equation}\label{9}
  2\left(\frac{\ddot{a}}{a}\right)+\left(\frac{\dot{a}}{a}\right)^2=0
\end{equation}
\begin{equation}\label{10}
  \dot{\rho}_{DM}+3\left(\frac{\dot{a}}{a}\right)\rho_{DM}=0
\end{equation}
 From \eqref{8},\eqref{9},\eqref{10} we obtain
 \begin{equation}\label{}
   a=a_0\left(\frac{t}{t_0}\right)^{\frac{2}{3}},
 \end{equation}
 \begin{equation}\label{}
   \rho_{DM}=\rho_0\left(\frac{t_0}{t}\right)^{2}
 \end{equation}
 and the deceleration parameter
 \begin{equation}\label{}
   q = \frac{1}{2}.
 \end{equation}
This positive deceleration parameter indicates that the universe is experiencing a deceleration phase. However, type Ia supernovae observational data suggest that the universe is undergoing an accelerated expansion phase. Hence, we may conclude that our present universe cannot be filled with dark matter completely. \\
\\
\textbf{Case-II:} Perfect Fluid ($p_{PF}=\omega\rho_{PF}$)
\\
\\
Let us consider the universe  completely filled by a perfect fluid. The  field equations are:

\begin{equation}\label{14}
  3\left(\frac{\dot{a}}{a}\right)^2=\rho_{PF}
\end{equation}
\begin{equation}\label{15}
  2\left(\frac{\ddot{a}}{a}\right)+\left(\frac{\dot{a}}{a}\right)^2= - p_{PF}
\end{equation}
\begin{equation}\label{16}
  \dot{\rho}_{PF}+3\left(\frac{\dot{a}}{a}\right)(\rho_{PF} + p_{PF})=0
 \end{equation}
From \eqref{14} \eqref{15},\eqref{16}, we obtain
 \begin{equation}\label{}
   \rho_{PF}=\frac{\rho_0}{a^{3(1+\omega)}}
 \end{equation}
 and deceleration parameter
 \begin{equation}\label{}
   q = \frac{1+3\omega}{2}.
 \end{equation}
Here we can conclude that $q > 0$ for $0<\omega\le 1$ and similarly we cannot have accelerated expansion of the universe, when the universe is completely filled with perfect fluid.
Using velocity equation \eqref{14} and the acceleration equation \eqref{15}, we  find

\begin{equation}\label{19}
\frac{\ddot{a}}{a} = - \frac{(\rho + 3p)}{6}
\end{equation}
which is sometimes called the Raychaudhari equation. Note that here the acceleration and the deceleration of the universe depend on the sign of $\ddot{a}$. Hence, from equation \eqref{19}, we can say that the universe is decelerating if $\rho + 3p > 0$, while it is accelerating if $\rho + 3p < 0$. If the linear equation of state $p=\omega\rho$ holds, then the condition can be transferred to equation of state parameter with $\omega > - \frac{1}{3}$ for decelerating universe and $\omega < - \frac{1}{3}$ for accelerating universe. In this case, we are out of the Zeldovich interval.\\
\\
\textbf{Case-III:} Mixture of dark matter and perfect fluid
\\
\\
In this case, we consider the universe  filled with dark matter (the density is $\rho_{DM}$ and the pressure is equal to zero) and perfect fluid (with density $\rho_{PF}$ and pressure $p_{PF} = \omega \rho_{PF}$). We assume no interaction between dark matter and perfect fluid. The  field equations are
\begin{equation}\label{20}
  3\left(\frac{\dot{a}}{a}\right)^2=\rho_{PF} + \rho_{DM}
\end{equation}
\begin{equation}\label{21}
  2\left(\frac{\ddot{a}}{a}\right)+\left(\frac{\dot{a}}{a}\right)^2= - p_{PF}
\end{equation}
\begin{equation}\label{22}
  \dot{\rho}_{DM}+3\left(\frac{\dot{a}}{a}\right)\rho_{DM}=0
 \end{equation}
 \begin{equation}\label{23}
 \dot{\rho}_{PF}+3\left(\frac{\dot{a}}{a}\right)(\rho_{PF}+ p_{PF})=0
 \end{equation}
The density parameters are
\begin{equation}
\Omega_{DM} = \frac{\rho_{DM}}{3H^2} , ~~ \Omega_{PF} = \frac{\rho_{PF}}{3H^2} ,
\end{equation}
where $\displaystyle H=\left(\frac{\dot{a}}{a}\right)$.
In order to  introduce dynamical systems,  we define
\begin{equation}
x = \frac{\rho_{DM}}{3H^2} ~~ \text{and} ~~ 1-x = \frac{\rho_{PF}}{3H^2}\,,
\end{equation}
according to the fact that the density constraint is $\Omega_{DM} + \Omega_{PF} = 1 $. These assumptions  lead us to the following  autonomous equation
\begin{equation}
x' = f(x) = 3 \omega x(1 - x)
\end{equation}
The combination of various fluids is represented in Fig.(1).

\begin{figure}
\label{case3_2}
\includegraphics[scale=1]{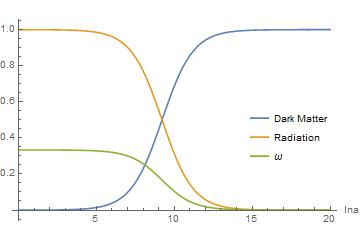}
\caption{The qualitative evolutions of various fluids.}
\end{figure}
where prime denotes derivative with respect to $d \eta = Hdt$ and the variable $x$ is a function of the time parameter $\eta = \ln a$. There are only two critical points for this dynamical system, i.e. $x=0$ and $x=1$. The first one is unstable ($f'(0) > 0$), while the second one is stable ($f'(1) < 0$). The evolution of the universe starts from the completely radiation dominated universe and ends up in a matter dominated era. The corresponding equation of state for this universe is ${\displaystyle \omega_{eff} = \omega_{DM}\Omega_{DM} + \omega_{R}\Omega_{R} = \frac{1-x}{3}} $, which starts from the value of $\displaystyle \frac{1}{3}$ and then drops to zero as matter starts to dominate.

\begin{center}
\begin{tabular}{ |c|c|c| }
 \hline \textbf{Critical Point} & \textbf{Stability} \\
 \hline
 $0$ & Unstable \\
 \hline
 $1$ & Stable \\
 \hline
\end{tabular}
\end{center}
\textbf{Case-IV:} Mixture of dark matter, perfect fluid, and dark energy.
\\
\\
Let us consider the universe filled with perfect fluid (with density $\rho_{PF}$ and pressure $p_{PF}$), non-relativistic dark matter (with density $\rho_{DM}$ and pressure $p$= 0) and dark energy (with density $\rho_{DE}$ and pressure $p_{DE}$). We assume that there is no interaction between dark energy and dark matter. The  field equations are:

\begin{equation}
3\left(\frac{\dot{a}}{a}\right)^2=\rho_{PF} + \rho_{DM} + \rho_{DE}
\end{equation}

\begin{equation}
2\left(\frac{\ddot{a}}{a}\right)+\left(\frac{\dot{a}}{a}\right)^2= - (p_{PF}+p_{DE})
\end{equation}

\begin{equation}
\dot{\rho}_{DM} = -3H \rho_{DM}
\end{equation}

\begin{equation}
\dot{\rho}_{PF} = -3H \rho_{PF}(1+w)
\end{equation}

\begin{equation}
\dot{\rho}_{DE} = -3H \rho_{DE}(1+w_{DE})
\end{equation}
We can define the various density parameters as:
\begin{equation}
\Omega_{DM} = \frac{\rho_{DM}}{3H^2} , ~~ \Omega_{PF} = \frac{\rho_{PF}}{3H^2} , ~~ \Omega_{DE} = \frac{\rho_{DE}}{3H^2}\,,
\end{equation}
and  introduce the following dimensionless variables:
\begin{equation}
x = \frac{\rho_{DM}}{3H^2} , y = \frac{\rho_{PF}}{3H^2}
\end{equation}
So we have,
\begin{equation}
\Omega_{DE} = 1 - x - y\,.
\end{equation}
Assuming that dark matter, perfect fluid and dark energy do not interact  each other leads to the following set of autonomous differential equations:

\begin{equation}{\label{35}}
x' = 3x(\omega y  + \omega_{DE}(1-x-y))\,,
\end{equation}

\begin{equation}{\label{36}}
y' = 3y [\omega(y - 1) +  \omega_{DE}(1-x-y)]\,,
\end{equation}
where $O(0,0)$, $A(0,1)$, and $B(1,0)$ are the critical points of this system. Physically $(0,0)$  means that the universe is completely filled with dark energy, $(1,0)$ means that the universe is completely filled with dark matter and $(0,1)$  means that the universe is completely filled with perfect fluid.
The Jacobian Matrix for this set of autonomous equations is:
\[\begin{bmatrix}
3 (\omega_{DE} (1-x-y) +  \omega y)-3 \omega_{DE} x & 3 (\omega -\omega_{DE}) x \\
 -3 \omega_{DE} y & 3 (\omega_{DE} (1-x-y) +\omega (y-1) +3 (\omega - \omega_{DE}) y \\
\end{bmatrix}\]
By evaluating the Jacobian at the above mentioned critical points and finding its eigenvalues, we get:

\begin{center}
\begin{tabular}{ |c|c|c|c| }
 \hline \textbf{Point} &  \textbf{$\omega_{eff}$} & \textbf{Eigenvalues} & \textbf{Stability} \\
 \hline
 (0, 0) &  $\omega_{DE}$ & 3$\omega_{DE}$, $3( -\omega + \omega_{DE})$ & Stable\\
 \hline
 (0, 1) &  $\omega$ & 3$\omega$, $3(\omega - \omega_{DE})$ & Unstable \\
 \hline
  (1, 0) &  $0$ & $-3\omega_{DE}$, $-3 \omega$ & Saddle Point \\
 \hline
\end{tabular}
\end{center}
Considering radiation as a particular case of the perfect fluid (${\displaystyle \omega_{R} = \frac{1}{3}}$),  we  have:

\begin{center}
\begin{tabular}{ |c|c|c|c| }
 \hline \textbf{Point} &  \textbf{$\omega_{eff}$} & \textbf{Eigenvalues} & \textbf{Stability} \\
 \hline
 (0, 0) &  $\omega_{DE}$ & 3$\omega_{DE}$, $3 \omega_{DE} -1$ & Stable\\
 \hline
 (0, 1) &  $\frac{1}{3}$ & $1$, 1- $3(\omega_{DE})$ & Unstable \\
 \hline
 (1, 0) &  $0$ & $-3\omega_{DE}$, $-1$ & Saddle Point \\
 \hline
\end{tabular}
\end{center}
\textbf{Subcase I}:\\
Furthermore, if we consider cosmological constant as a particular type of dark energy $(w_{\Lambda} = -1)$, we  have:

\begin{center}
\begin{tabular}{ |c|c|c|c| }
 \hline \textbf{Point} &  \textbf{$\omega_{eff}$} & \textbf{Eigenvalues} & \textbf{Stability} \\
 \hline
 (0, 0) &  $-1$ & {-3,-4} & Stable\\
 \hline
 (0, 1) &  $\frac{1}{3}$ & {1,4} & Unstable \\
 \hline
 (1, 0) &  $0$ & {3,-1} & Saddle Point \\
 \hline
\end{tabular}
\end{center}
For similar results see also \cite{Tamanini}.
Considering equations \eqref{35} and \eqref{36} with the above assumption, we get the following set of autonomous  equations

\begin{equation}{\label{37}}
x' = x(3x + 4y -3)
\end{equation}

\begin{equation}{\label{38}}
y' = y(3x + 4y -4)
\end{equation}
The phase space portrait of the dynamical system \eqref{37} and \eqref{38} is plotted in Fig (2). Clearly origin is the stable point in this system.
\begin{figure}
\includegraphics[scale=0.4]{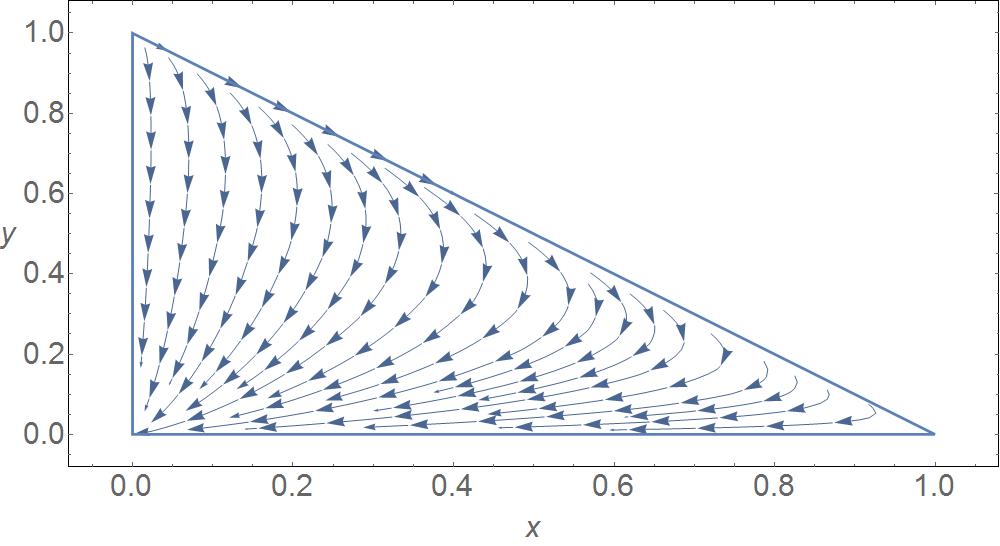}
\caption{The phase space portrait the dynamical system \eqref{37} and \eqref{38}.}
\end{figure}
We can also study the relative energy density of dark matter, radiation and dark energy together with effective equation of state parameter in $\Lambda CDM$ model.
\begin{figure}
\includegraphics[scale=0.4]{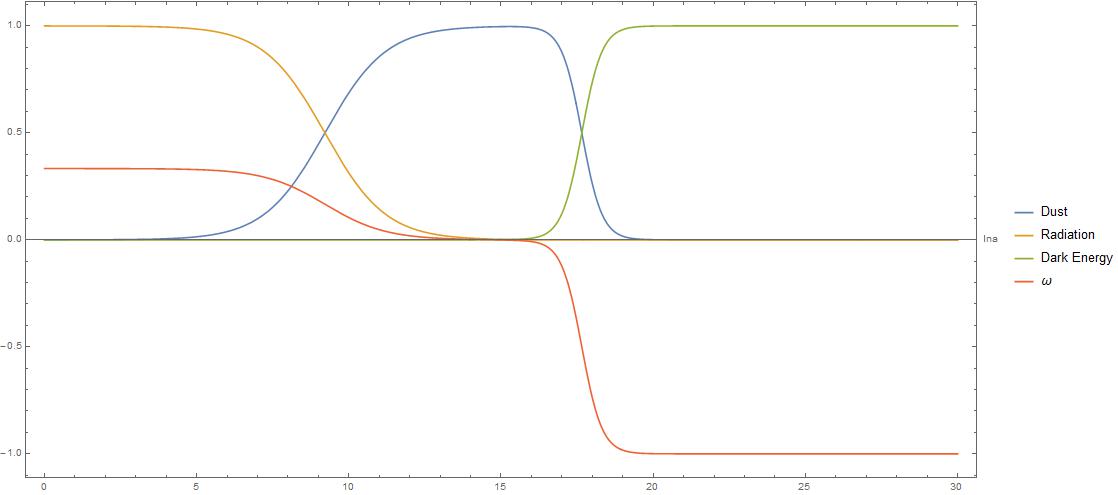}
\caption{The qualitative evolution of dark matter, radiation, and dark energy.}
\end{figure}
In the above picture, we see that during initial stage, we have universe completely filled with radiation, which then slowly reduces and matter starts to form, hence increasing its relative energy density. In the later stage, dark energy starts to dominate and hence causes the accelerated expansion of the universe which we currently observe. The equation of state parameter $\omega_{eff}$ starts with $\frac{1}{3}$ when  the universe is completely radiation dominated then during the matter formation era, it goes down to zero and further, when dark energy starts to dominates, it becomes negative and ultimately reaches the value of $\omega_{\Lambda}$ which is -1. \\
\textbf{Subcase II}:\\
Now we assume stiff fluid as a form of the perfect fluid and we still stick to cosmological constant. Here we get a different set of autonomous differential equations which are

\begin{equation}{\label{39}}
x' = 3x(x+2y -1)
\end{equation}

\begin{equation}{\label{40}}
y' = 3y(x+2y-2)
\end{equation}
Here, the corresponding set of critical points and stability conditions remains same, but as the system of differential equations is different, so we get a different set of eigenvalues, which is as follows:

\begin{center}
\begin{tabular}{ |c|c|c|c| }
 \hline \textbf{Point} &  \textbf{$\omega_{eff}$} & \textbf{Eigenvalues} & \textbf{Stability} \\
 \hline
 (0, 0) &  $-1$ & {-3,-6} & Stable\\
 \hline
 (0, 1) &  $\frac{1}{3}$ & {3,6} & Unstable \\
 \hline
 (1, 0) &  $0$ & {3,-3} & Saddle Point \\
 \hline
\end{tabular}
\end{center}
For the dynamical set of equations \eqref{39} and \eqref{40}, again origin is the critical point here and we have a phase space portrait as given in Fig.(4).
\begin{figure}
\includegraphics[scale=0.4]{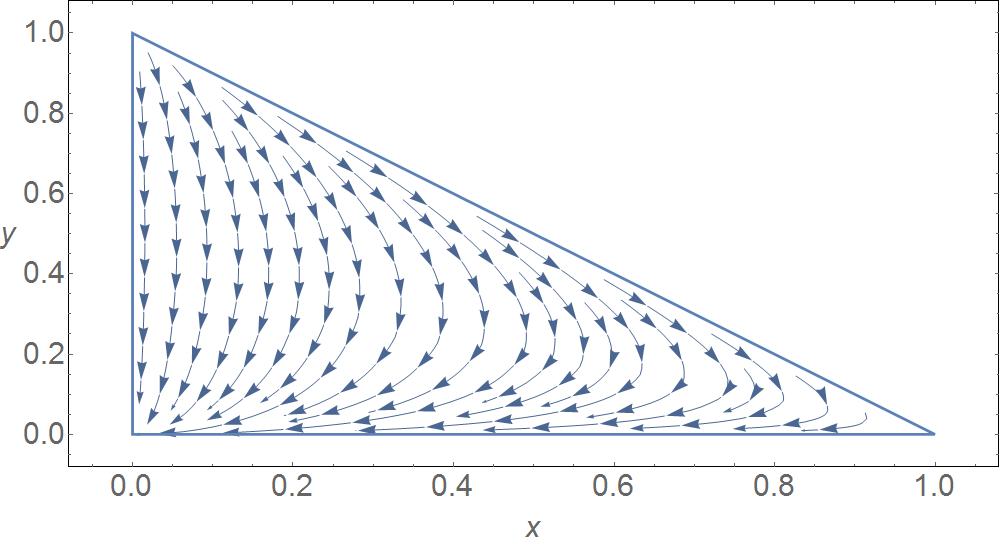}
\caption{The phase space portrait of system \eqref{39}, \eqref{40}.}
\end{figure}
Here also we can study the relative energy density of dark matter, stiff fluid and dark energy together with effective equation of state parameter in $\Lambda CDM$ model. \\
\begin{figure}
\includegraphics[scale=0.4]{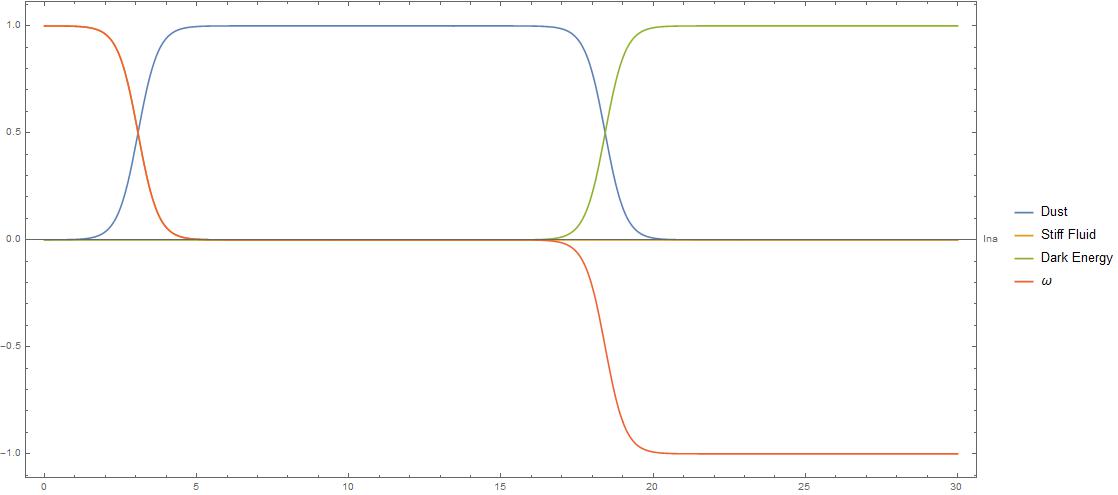}
\caption{The qualitative evolution of dark matter, stiff matter, and dark energy.}
\end{figure}
In Fig. (5),  we see that during initial stage, we have universe completely filled with stiff fluid, which then slowly reduces and matter starts to form, hence increasing its relative energy density. So $\omega_eff$ is behaving same as energy density of stiff fluid, till dark energy comes into the picture.  In the later stage,  dark energy starts to dominate and causes the accelerated expansion of the universe which we currently observe. The parameter $\omega_{eff}$ starts with $\frac{1}{3}$ as universe was completely radiation dominated then during the matter formation era, it goes down to zero and further when dark energy starts to dominates, it goes negative and ultimately reach the value of $\omega_{\Lambda}$ which is -1. \\
\textbf{Subcase III}:\\
As there was no significant difference in this phase portrait, so we now  analyze this system by taking $\omega_{DE} \to -\infty$. As discussed previously, the stability conditions are independent of values of equation of state parameter $(\omega)$. But stability comes faster and prominent with $\omega_{DE} \to -\infty$. Here it can be seen that when there was cosmological constant alongside radiation and dark matter, the universe was moving from radiation domination towards dark matter domination which is the saddle point and then ultimately moving towards the stable point (0,0). But here, whatever may be the beginning, the universe quickly moves towards the stable point (0,0) and hence makes the saddle point (1,0) weak.  Conversely, if we take $\omega_{DE} \to 0$, but negative, then we observe the universe to  move from radiation domination to dark matter domination and then to dark energy domination quite slowly, which makes the saddle point (1,0) stronger and the stable point (0,0) weaker. See Figs. (6) and (7).
\begin{figure}
\includegraphics[scale=0.4]{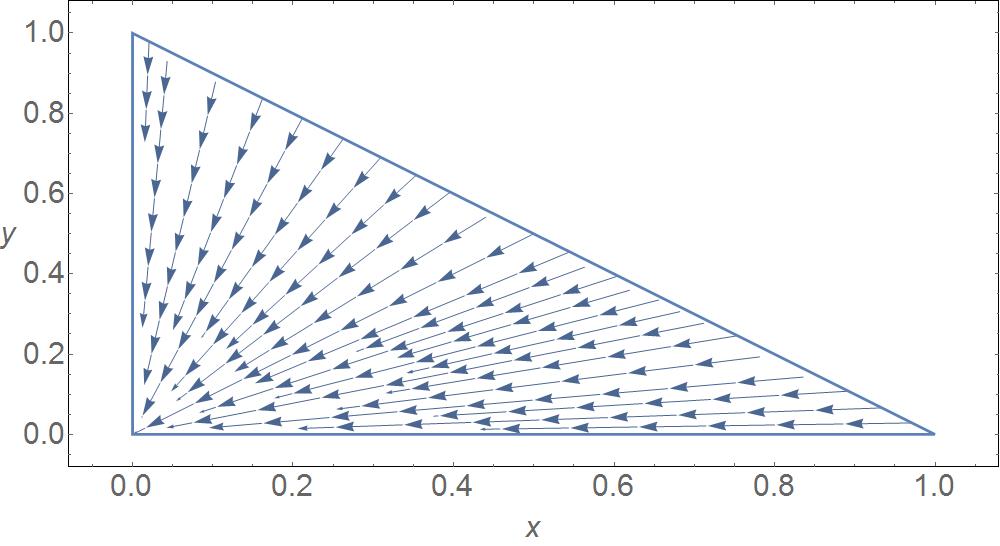}
\caption{The phase space portrait of stable  point (0,0).}
\end{figure}
\begin{figure}
\includegraphics[scale=0.4]{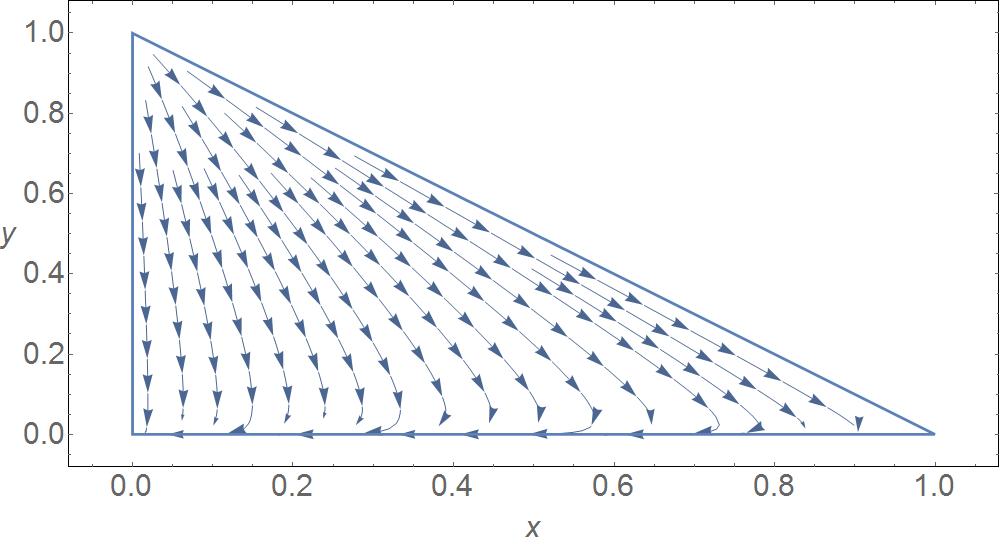}
\caption{The phase space portrait of saddle point (1,0).}
\end{figure}

So from this case-IV, we may conclude one important fact that the critical points are not $\omega$ dependent quantities for the mixture of dark matter, perfect fluid and dark energy. We get a different set of eigenvalues for different kinds of mixture, which leads to some difference in stability properties which have been discussed.
\\ \\
\textbf{Case V} Mixture of dark matter, Perfect fluid and scalar field as a form of dark energy.
\\
Let us consider the universe filled with perfect fluid (with density $\rho_{PF}$ and pressure $p_{PF}$), non-relativistic dark matter (with pressure equal to zero and  density $\rho_{DM}$), and scalar field (as a form of dark energy). We assume that there is no interaction between components. The  field equations are:

\begin{equation}{\label{41}}
3\frac{\dot{a}^2}{a^2} = \rho_{dm} + \rho_{pf} + \frac{1}{2} \dot{\phi}^2 + V(\phi)
\end{equation}

\begin{equation}{\label{42}}
2\frac{\ddot{a}}{a}+\frac{\dot{a}^2}{a^2} = -p_{pf} - \frac{1}{2} \dot{\phi}^2 + V(\phi)
\end{equation}

\begin{equation}{\label{43}}
\dot{\rho}_{DM} = -3H \rho_{DM}
\end{equation}

\begin{equation}{\label{44}}
\dot{\rho}_{PF} = -3H \rho_{PF}(1+w)
\end{equation}

\begin{equation}{\label{45}}
\dot{\rho}_{\phi} = -3H \rho_{\phi}(1+w_{\phi})
\end{equation}
Let us introduce dimensionless quantities as which are  referred as Expansion Normalized variables (EN) \cite{Wainwright}:
\begin{equation}
x = \frac{\dot{\phi}}{\sqrt{6}H} , y = \frac{\sqrt{V}}{\sqrt{3}H}, z = \frac{\sqrt{\rho_{ dm}}}{\sqrt{3}H}\,.
\end{equation}
With these variables and the Friedman constraint, we have,
\begin{equation}
\Omega_{pf} + \Omega_{dm} + \Omega_{\phi} = \Omega_{pf} + x^2 + y^2 + z^2 = 1
\end{equation}
where relative energy densities are $\Omega_{dm} =\frac{\rho_{dm}}{3H^2}$ and $\Omega_{\phi} = \frac{\rho_{\phi}}{3H^2}$.
This effectively gives us
\begin{equation}
\Omega_{pf} = 1 - (x^2 + y^2 + z^2)\,.
\end{equation}
With the realistic assumption  $\rho \geq 0$, we have $x^2 + y^2 + z^2 \leq 1$.
One can  also define
\begin{equation}
\lambda = - \frac{V_{,\phi}}{V}\,.
\end{equation}
Here we note that these EN variables fail to be a system of   autonomous equations as $\lambda$ still depends on the scalar field $\phi$. We can consider an  exponential potential, where $\lambda$ is just a parameter and system of equations becomes autonomous. In general, the equation for variable $\lambda$ is given by:
\begin{equation}
\lambda' = \frac{1}{H}\left(\frac{d\lambda}{d \phi}\right)\,,
\end{equation}
so we  have,
\begin{equation}
\lambda' = - \sqrt{6}(\Gamma - 1)\lambda^2 x
\end{equation}
where, $\displaystyle \Gamma = \frac{V V_{,\phi \phi}}{V_{\phi}^2}$.

\textbf{Subcase I}: Exponential Potential \\
We assume an exponential potential  $V=V_0 e^{-\lambda\phi}$ with  $\Gamma = 1$, hence $\lambda$ is just a parameter, which  means that y $>$ 0 and our phase space becomes the upper half of the sphere.
Dynamical systems for cosmological scalar fields with an exponential potential has been widely  studied in view of cosmic acceleration\cite{Copeland1}. All assumptions considered above  lead  to the  following system of autonomous equations
\begin{eqnarray} \label{52}
x' &=& - \frac{3}{2} \left[ 2x + (\omega -1)x^3 + x( \omega + 1)(y^2-1) + x \omega z^2 - \frac{\sqrt{2}}{\sqrt{3}} \lambda y^2 \right] \\ \label{53}
y' &=& - \frac{3}{2}y \left[ (\omega -1) x^2 + (\omega + 1)(y^2 -1) + \omega z^2 + \frac{\sqrt{2}}{\sqrt{3}} \lambda x \right] \\ \label{54}
z' &=& -\frac{3}{2}z \left[ -1 + (\omega -1)x^2 + (w+1)(y^2 - 1) + \omega z^2 \right].
\end{eqnarray}
System of equations \eqref{52}, \eqref{53}, \eqref{54} remains invariant under the transformation y $\rightarrow$ -y and z $\rightarrow$ -z.
Now the acceleration equation gives
\begin{equation}
\frac{\dot{H}}{H^2} = \frac{3}{2} \left [ (\omega -1)x^2 + (\omega + 1)(y^2 - 1) + \omega z^2 \right].
\end{equation}
We define the effective equation of state parameter of the universe as $\displaystyle \omega_{eff} = \frac{(p_{pf} + p_{\phi})}{(\rho_{dm} + \rho_{pf} + \rho_{\phi})} $ which can be written in terms of EN variables as
\begin{equation}
\omega_{eff} = x^2_{*} - y^2_{*} + \omega (1 - x^2_{*} - y^2_{*} - z^2_{*}),
\end{equation}
where $(x_*,y_*,z_*)$ is any point in our phase space and, for $\displaystyle \omega_{eff} < - \frac{1}{3}$, it describes a universe undergoing accelerated phase of expansion.

The Jacobian of the set of autonomous differential equations is:

\[\begin{bmatrix}
3 (\omega-1) x^2+w z^2+(\omega+1)(y^2-1)+2 & 2(\omega + 1) x y-2 \sqrt{\frac{2}{3}} y \lambda & 2 \omega x z \\
 -\frac{3}{2} y (2(\omega-1) x + \sqrt{\frac{2}{3}} \lambda) &
\begin{split}
 -3 (\omega+1) y^2-\frac{3}{2} (\omega-1) x^2 \\ +\sqrt{\frac{2}{3}} \lambda  x+w z^2+(w+1)(y^2-1))
 \end{split}
 & -3 \omega y z \\
-3 (\omega-1) x z & -3 (\omega + 1) y z &
\begin{split}
-3 \omega z^2 - \frac{3}{2} ((\omega - 1) x^2 \\ + \omega z^2 +(\omega+1)(y^2-1)-1)
\end{split}
\\

\end{bmatrix}\]
Real critical points include
$(0,0,0), (\pm 1,0,0), \left(0,0,\pm \frac{\sqrt{2+\omega}}{\sqrt{\omega}}\right),\left(\frac{\sqrt{\frac{3}{2}}(1+ \omega)}{\lambda},\pm \frac{\sqrt{\frac{3}{2}(1- \omega^2)}}{\lambda},0\right)$ and  $\left(\frac{\lambda}{\sqrt{6}},\pm \sqrt{\frac{6 - \lambda^2}{6}},0\right)$.
Let us now study the behavior of each of these points in detail.
\\ \\
Point $A \equiv(0,0,0)$ behaves as a saddle point as eigenvalues take both the sign. Also $\omega_{eff}$ in this region is $\omega$, which is in the interval [0,1]. So there cannot be any accelerated region for this point. In this case all the values i.e. $x, y, z$ are zero, hence this type of universe is a perfect fluid dominated.
\\ \\
Point $B_1\equiv (1,0,0)$ is unstable for $\lambda \leq \sqrt{6}$ and saddle point for $\lambda > \sqrt{6}$. Also $\omega_{eff}$ in this region is 1. As  it is  $x=1$, this model of universe is completely scalar field dominated. Also in this case, we cannot have accelerated phase for this model.
\\ \\
Point $B_2\equiv (-1,0,0)$ is unstable for $\lambda \geq -\sqrt{6}$ and saddle point for $\lambda < -\sqrt{6}$. Also $\omega_{eff}$ in this region is 1. Like in the previous case, this is a completely scalar field dominated system and we cannot have accelerated phase for this model.
\\ \\
Point $C\equiv\left(0,0,\pm \frac{\sqrt{2+\omega}}{\sqrt{\omega}}\right)$ is not a useful critical point  because the condition of $x^2 + y^2 + z^2 \leq 1$ is not  satisfied. So no further evaluation regarding this critical point can be considered.
\\ \\
Point $D\equiv\left(\frac{\lambda}{\sqrt{6}}, \pm \sqrt{\frac{6 - \lambda^2}{6}}, 0\right)$ behaves as a saddle point if $\lambda^2 \leq 6$, else it becomes unstable. The parameter $\omega_{eff}$ here becomes $\frac{\lambda^2}{3} -1 $, which, for $\lambda^2 < 2 $,  gives an  accelerated phase. This model is scalar field dominated.
\\ \\
Point $E\equiv\left(\frac{\sqrt{\frac{3}{2}}(1+ \omega)}{\lambda},\pm \frac{\sqrt{\frac{3}{2}(1- \omega^2)}}{\lambda}, 0\right)$ cannot be stable as we have one eigenvalue for $\frac{3(2+\omega)}{2}$ which is positive. Apart from this, this point  exists for $\lambda^2 \geq 3(1+ \omega)$ and we cannot have accelerated phase because $\omega_{eff} = \omega$ which is in the interval  [0,1]. \\
\\
A detailed description of points with their eigenvalues and stability is given in the following Tables.
\begin{center}
\begin{tabular}{ |c|c|c| }
 \hline \textbf{Critical Point} & \textbf{Eigenvalues} & \textbf{Stability} \\
 \hline
 $(0,0,0)$ & - $\frac{3}{2}(\pm 1 -\omega), \frac{3}{2}( 2 + \omega)$ & Saddle Point \\
 \hline
 $(1,0,0)$ & -$3(\omega -1), -\frac{3}{2}(-2 + \sqrt{\frac{2}{3}}\lambda), \frac{9}{2}$ & Unstable for $\lambda \leq \sqrt{6}$ \\ & & Saddle for $\lambda > \sqrt{6}$ \\
 \hline
 $(-1,0,0)$ & -$3(\omega -1), -\frac{3}{2}(-2 - \sqrt{\frac{2}{3}}\lambda), \frac{9}{2}$ & Unstable for $\lambda \geq - \sqrt{6}$ \\ & & Saddle for $\lambda < - \sqrt{6}$ \\
 \hline
 $\left(\frac{\lambda}{\sqrt{6}}, \pm \sqrt{\frac{6 - \lambda^2}{6}}, 0\right)$ & $\frac{1}{2} (\lambda^2 - 6), \frac{1}{2} (3 + \lambda^2), \lambda^2 - 3 \omega -3$ & Saddle Point if $\lambda^2 < 6$, \\ & & else Unstable \\
 \hline

\end{tabular}
\end{center}


\begin{center}
\begin{tabular}{ |c|c|c|c| }
 \hline
 \textbf{Critical Point} & \textbf{$\omega_{eff}$} & \textbf{Acceleration Phase} & \textbf{Existence} \\
 \hline
 $(0,0,0)$ & $\omega$ & No & all $\lambda$ and all $\omega$ \\
 \hline
 $(1,0,0)$ and $(-1,0,0)$ & 1 & No & all $\lambda$ and all $\omega$ \\
 \hline
 $(\frac{\lambda}{\sqrt{6}}, \pm \sqrt{\frac{6 - \lambda^2}{6}}, 0)$ & $\frac{\lambda^{2}}{3} - 1$ & $\lambda^{2} < 2$ & $\lambda^2 < 6$ \\
 \hline

\end{tabular}
\end{center}

\textbf{Subcase II}: Power Law Potential\\
Here we  consider power law type potential. Explicitly we have:
\begin{equation}
V(\phi) = \frac{M^{\alpha+4}}{\phi^{\alpha}}
\end{equation}
where $\alpha$ is a dimensionless parameter and $M$ is a positive constant with  mass dimensions.
For this potential, we have
\begin{equation}
\Gamma = \frac{V V_{,\phi \phi}}{V_{\phi}^2} = \frac{\alpha +1}{\alpha}
\end{equation}
or
\begin{equation}
\Gamma -1 = \frac{1}{\alpha}
\end{equation}
As $\Gamma = 1$, corresponds to exponential potential studied in Subcase I, in which $\lambda$ becomes a constant, we will exclude that case from here.
We can now here define a new variable $u$ as
\begin{equation}{\label{60}}
u = \frac{\lambda}{\lambda + 1},
\end{equation}
when $\lambda =0$, we get again $u =0$, but when $\lambda\rightarrow + \infty$, we have $u =1$, meaning that this new variable is only bounded to $0 \leq u \leq 1$.
In the new variable \eqref{60}, the dynamical system becomes:
\begin{eqnarray}
x' &=& - \frac{3}{2} \left[ 2x + (\omega -1)x^3 + x( \omega + 1)(y^2-1) + x \omega z^2 - \frac{\sqrt{2}}{\sqrt{3}} y^2 \frac{u}{1-u} \right] \\
y' &=& - \frac{3}{2}y \left[ (\omega -1) x^2 + (\omega + 1)(y^2 -1) + \omega z^2 + \frac{\sqrt{2}}{\sqrt{3}} x \frac{u}{1-u} \right] \\
z' &=& -\frac{3}{2}z \left[ -1 + (\omega -1)x^2 + (w+1)(y^2 - 1) + \omega z^2 \right] \\
u' &=& - \sqrt{6}(\Gamma -1)x u^2
\end{eqnarray}
The Jacobian matrix for this system of equation is
\[\begin{bmatrix}
\begin{split}
- \frac{9}{2} \left(\omega - 1 \right) x^2 + \omega z^2 + \\ \left( \omega + 1 \right) \left( y^2 -1 \right) + 2
\end{split}
& -\frac{3}{2} \left( 2 \left( \omega +1 \right) xy + \frac{2 \sqrt{\frac{2}{3}} u y}{1-u} \right) & -3 \omega x z & -\frac{3}{2}\left(-\frac{\sqrt{\frac{2}{3}} y^2}{1-u}-\frac{\sqrt{\frac{2}{3}} u y^2}{(1-u)^2}\right).\\
-\frac{3}{2} y \left(\frac{\sqrt{\frac{2}{3}} u}{1-u}+2 (w-1) x\right)&
\begin{split}
-\frac{3}{2} \left(\frac{\sqrt{\frac{2}{3}} u x}{1-u}+(\omega-1) x^2\right)\\ -\frac{3}{2} ((\omega+1) (y^2-1) \\+\omega z^2)-3 (\omega+1) y^2)
\end{split}
&-3 \omega y z& - \frac{3}{2}y \left(\frac{\sqrt{\frac{2}{3}} u x}{(1-u)^2}+\frac{\sqrt{\frac{2}{3}} x}{1-u}\right)\\
2 (\omega -1) x z & 2 (\omega +1) y z &
\begin{split}
(\omega -1) x^2+\\(\omega+1)(y^2-1) \\ +3 \omega z^2-1
\end{split}
 & 0 \\
\left(-\sqrt{6} (t-1) u^2\right)
 &0 & 0 & -2 \sqrt{6} (t-1) u x
\end{bmatrix}\]
By substituting critical points in this Jacobian Matrix, we can study detailed description of point with its eigenvalues and stability as given in the Table below:
\begin{center}
\begin{tabular}{ |c|c|c|c| }
 \hline \textbf{Critical Point} & \textbf{Eigenvalues} & \textbf{Hyperbolicity}& \textbf{Stability} \\
 \hline
 $(0,0,0,u)$ & $0, \frac{3}{2}(1 -\omega), -\frac{3}{2}( 1 - \omega), -2 - \omega $ & Non Hyperbolic & Saddle Point \\& & &\\
 \hline
 $(\pm 1,0,0,0)$ & 0, $3 (\omega -1), 3 , -3 $ & Non Hyperbolic & Saddle Point \\
 \hline
 $(0,1,0,0)$ & 0, -3, -1, $-3 (1 + \omega)$ & Non Hyperbolic & Stable Point for $\alpha > 0$ \\
 \hline
 $(0,-1,0,0)$ & 0, 3, -1, $-3 (1 + \omega)$ & Non Hyperbolic & Saddle Point \\
 \hline

\end{tabular}
\end{center}
Point $O\equiv(0,0,0,u)$ is the entire $u$ axis which is critical. Since these are not isolated critical points, we expect that, at least one eigenvalue of the Jacobian, vanishes. This is evident from the above Table. From the opposite signs of other eigenvalues, its is clear that this critical point behaves as a saddle point. Also it is a non-hyperbolic point, as one of the eigenvalues, and it  vanishes.
\\ \\
Point $A_{\pm}\equiv(\pm 1,0,0,0)$ is a scalar field kinetic dominated solution. So here $\omega_{eff} = \omega_{\phi} = 1$, which means that it is  a stiff-fluid dominated universe.  Similar to previous case, here also we have opposite signs of the eigenvalues, apart from one vanishing eigenvalue. So, again, it is a non-hyperbolic, saddle kind critical point.
\\ \\
Point $B\equiv (0,1,0,0)$ is a scalar field dominated point. Here $\omega_{eff} = \omega_{\phi} = -1$, which means that it is   dominated by the potential energy of the scalar field. Here, apart from one vanishing eigenvalue, all other eigenvalues are negative which leads to stability in the neighborhood of this point. Non-hyperbolicity still holds here due to a vanishing eigenvalue.
\\ \\
Point $C\equiv (0,-1,0,0)$ is also a scalar field dominated point. Here $\omega_{eff} = \omega_{\phi} = -1$, which means that it is  dominated by the potential energy of the scalar field. Here, apart from a vanishing eigenvalue, other eigenvalues have opposite signs which leads to a saddle point. Non-hyperbolicity still holds here due to a vanishing eigenvalue, which is evident from the Table above.

\section{Conclusion and Perspectives}
In this work, we have characterized the  cosmological evolution by considering a mixture of various fluids.
Precisely we considered five cases of fluid mixtures. We began  an entirely dark matter filled universe and derived its field equations and scale factor. The  deceleration parameter is  a positive constant, $\displaystyle \frac{1}{2}$ and then the model  is currently experiencing a deceleration phase, which contradicts the observational fact of accelerated expanding universe. We can  conclude that the universe cannot be completely dark matter dominated. Secondly, we consider a completely perfect fluid matter dominated universe with equation of state  $p=\omega \rho$ , with $0 < \omega \leq 1$. From similar calculations as in the previous case and, from Raychaudhari equation, it can be found that, for the accelerated expansion of the universe, $\omega$ must be less that $-\frac{1}{3}$. So it leads us to the conclusion that, for an accelerated expansion of the universe, it cannot  be completely perfect fluid filled. As a third case, we consider the mixture of both dark matter and perfect fluid, with both not interacting between each other. Then considering radiation as a kind of perfect fluid and  using dynamical system analysis, we obtain two critical points and conclude that such a universe will start with radiation domination which is an unstable point and slowly it reaches the matter dominated era, which is a stable point. Still in this case,  we cannot arrive to the today observed  accelerated expansion because $\omega_{eff}$ is  non-negative.

In the fourth case, we consider the non-interacting mixture of dark matter, perfect fluid and dark energy. We derive  a set of two autonomous differential equations whose critical points are suitable for  stability analysis. As a subcase, radiation is once considered as a kind of perfect fluid and cosmological constant as a kind of dark energy. In the next case, we stick to cosmological constant but consider stiff fluid as a  perfect fluid. We conclude that in all  such cases, the completely dark energy dominated universe is stable and it  leads  to the accelerated expansion. Energy density parameter can be taken into account  and its behavior is plotted against $\ln a$ to understand the domination of this quantity in  different eras of the universe. Point $(0,0)$, the stable critical point of this system  corresponding to dark energy dominated universe, is considered  for various kinds of dark energies such as the cosmological constant, $\omega_{DE} \to -\infty$ and $\omega_{DE} \to 0$ (but negative). We conclude that the more stronger  dark  energy os (low value of $\omega_{DE}$), the  more stronger the stable critical point $(0,0)$ is. It is also evident that as $\omega_{DE} \to -\infty$, the critical point $(0,0)$ is so strong that there is no role for the saddle point in that case.

In the fifth case, we consider universe filled with a non-interacting mixture of dark matter, perfect fluid and scalar field with potential as a form of dark energy. Exponential potential has been considered in the first subcase. Here we derive a  set of three autonomous differential equations from which critical points are obtained and Jacobian matrix has been constructed. We discuss the eigenvalue stability analysis as  in the previous cases. The acceleration phase has to be  $\displaystyle \omega_{eff}<\frac{1}{3}$ to be achieved. Subsequently, we consider a power law potential and  we obtain a set of four autonomous differential equations, where the fourth equation  depends on the potential parameter $\lambda$. We arrive at the critical point which is a completely scalar field dominated being a stable point in this case.
The use of the dynamical system technique  allows to understand the behavior of various models by the analysis of $\omega_{eff}$. The values of the critical points in the phase space allows to determine the domination of the particular fluid. Then, using the critical point and evaluating Jacobian at those points, we arrive at the eigenvalues of the system. The signs of these eigenvalues allow to understand the stability of particular points. As a final remark, in this paper, we considered the simplest case of Einstein gravity and minimally coupled fluids.
More general cases and extensions can be considered in this perspective \cite{rept}.

\section{Acknowledgement}
The author Parth Shah is extremely thankful to Department of Science and Technology (DST), Govt. of India, for providing INSPIRE Fellowship (Ref. No. IF160358) for carrying out his research work. G. C. Samanta is thankful to Council of
Scientific and Industrial Research (CSIR), Govt. of India, for providing support (Ref.
No. 25(0260)/17/EMR-II) for carrying out the research work.
SC acknowledges INFN Sez. di Napoli (Iniziativa Specifica QGSKY).
This article is based upon work from COST Action (CA15117, CANTATA), supported by COST (European Cooperation in Science and Technology). Furthermore, the authors are very much thankful to the reviewers for their valuable comments for improvement of our work.


\begin{thebibliography}{}
\bibitem {riess} A. G. Riess et al., Astron. J. \textbf{116}(1998) 1009.

\bibitem {perlmutter} S. Perlmutter et al., Astrophys. J. \textbf{517} (1999) 565.

\bibitem {weinberg} S. Weinberg, Rev. Mod. Phys. \textbf{61} (1989) 1.

\bibitem {martin} J. Martin, C. R. Phys. \textbf{13} (2012) 566.

\bibitem {zlatev} I. Zlatev, L. M. Wang and P. J. Steinhardt, Phys. Rev. Lett. \textbf{82} (1999) 896.

\bibitem{Amendola} L. Amendola and S. Tsujikawa, Dark Energy: Theory and Observations, Cambridge Univ. Press, Cambridge (2010)

\bibitem {copeland} E. J. Copeland, M. Sami and S. Tsujikawa, Int. J. Mod. Phys. D \textbf{15} (2006) 1753.

\bibitem {tsujikawa} S. Tsujikawa, Class. Quanttum Grav. \textbf{30} (2013) 214003.

\bibitem{Arturo} L. Arturo et.al. , Journal of Physics: Conference Series \textbf{761} (2016) 012076.

\bibitem{Ellis} J. Wainright and G.F.R. Ellis, Dynamical Systems in Cosmology, Cambridge Univ. Press, Cambridge  (1997).

\bibitem{Coley} A. A. Coley, Dynamical Systems and Cosmology, Springer Ed.,  New York (2003).


\bibitem{Place} D.K. Arrowsmith, C.A. Place,  An Introduction to Dynamical Systems,
Cambridge University Press, Cambridge (1990).

\bibitem{Hirsch} M.W. Hirsch, R.L. Devaney and S. Smale, Differential Equations, Dynamical Systems and Introduction to Chaos, Academic Press, Elsevier, Amsterdam (1974).

\bibitem{Lefschetz} S. Lefschetz, Differential Equations. Geometric Theory, Interscience, New York (1957).

\bibitem{Lynch} S. Lynch, Dynamical Systems with applications using Mathematica, Springer Ed.,  Berlin (2007).

\bibitem{Perko} L. Perko, Differential equations and Dynamical systems, Springer Ed., Berlin  (2001).

\bibitem{Wiggins} S. Wiggins, Introduction to Applied Nonlinear Dynamical Systems and Chaos (2003).

\bibitem{Wainwright} J. Wainwright, G. F. R. Ellis, Dynamical Systems in Cosmology, Cambridge University Press (2005).

\bibitem{Copeland1} E. J. Copeland et. al.,Phys. Rev. D \textbf{57} (1998) 4686.

\bibitem{Kbamba} K. Bamba, S. Capozziello, S. Nojiri and S. D. Odintsov, Astrophys. and Space Sci. \textbf{342} (2012)  155.



\bibitem{Roy} N. Roy, arXiv: 1511.07978[gr-qc], 2015.

\bibitem{Bamba} K. Bamba,  D. Momeni and M. Al Ajmi,   arXiv: 1711.10475[gr-qc] (2017).

\bibitem{Odintsov} S. D. Odintsov and V. K. Oikonomou, Phys.Rev. D \textbf{96} (2017) 104049.

\bibitem{Odintsov1} S. D. Odintsov, V. K. Oikonomou and Petr V. Tretyakov, Phys.Rev. D \textbf{96} (2017) 044022.

\bibitem{Hohmann} M. Hohmann, L. Jarv and U. Ualikhanova, Phys.Rev. D \textbf{96} (2017) 043508.


\bibitem{Bhatia} A. S. Bhatia and S. Sur, Int. J. Mod. Phys. D \textbf{26} (2017) 1750149.

\bibitem{Carneiro} S. Carneiro and H Borges, Gen. Rel. Grav. \textbf{50} (2018) 1.

\bibitem{Santos}
  S.~Santos Da Costa, F.~V.~Roig, J.~S.~Alcaniz, S.~Capozziello, M.~De Laurentis and M.~Benetti,
    Class.\ Quant.\ Grav.\  {\bf 35} (2018),  075013
    

 \bibitem{Tamanini} N. Tamanini, Dynamics of cosmological scalar fields, Phys. Rev. D \textbf{89} (2014) 083521

\bibitem{rept}
  S.~Capozziello and M.~De Laurentis, Phys.\ Rept.\  {\bf 509} (2011) 167



\end{thebibliography}
\end{document}